\documentclass[prc,superscriptaddress,unsortedaddress,twocolumn,showpacs,
  floatfix,hyperref]{revtex4}
\usepackage[dvips]{graphicx}
\usepackage{amsmath}
\usepackage{bm}
\newcommand{\paral}{{\parallel}}
\newcommand{\vek}[1]{\bm{\mathrm{#1}}}
\newcommand{\eq}{\mathrm{eq}}
\begin{document}
\title{Collective excitations in the neutron star inner crust}
\author{Luc Di Gallo} \email{luc.digallo@obspm.fr} 
\author{Micaela Oertel} \email{micaela.oertel@obspm.fr} 
\affiliation{LUTH, Observatoire de Paris, CNRS, Universit\'e
  Paris Diderot, 5 place Jules Janssen, 92195 Meudon, France} 
\author{Michael Urban} \email{urban@ipno.in2p3.fr} 
\affiliation{Institut de Physique Nucl\'eaire,
  CNRS/IN2P3 and Universit\'e Paris Sud 11, 91406 Orsay, France}
\begin{abstract}
  We study the spectrum of collective excitations in the inhomogeneous
  phases in the neutron star inner crust within a superfluid
  hydrodynamics approach. Our aim is to describe the whole range of
  wavelengths, from the long-wavelength limit which can be described
  by macroscopic approaches and which is crucial for the low-energy
  part of the spectrum, to wavelengths of the order of the dimensions
  of the Wigner-Seitz cells, corresponding to the modes usually
  described in microscopic calculations. As an application, we will
  discuss the contribution of these collective modes to the specific
  heat of the ``lasagna'' phase in comparison with other known
  contributions.
\end{abstract}
\maketitle

\section{Introduction}
Neutron stars are fascinating objects, containing matter under extreme
conditions of temperature, density and magnetic field. In order to
study these celestial bodies, theoretical modeling has to be
confronted with observations. A prominent observable is the thermal
evolution of isolated neutron stars. Properties of the crust thereby
influence the cooling process mainly during the first 50-100 years,
when the crust stays hotter than the core which cools down very
efficiently via neutrino emission (see e.g. \cite{Yakovlev07}). Heat
transport in the crust is the key ingredient to explain the afterburst
relaxation in X-ray transients, too~\cite{Shternin07,Brown09}.
Concerning the models for the thermal relaxation of the crust, the
most important microscopic ingredients are thermal conductivity and
heat capacity, and to less extent neutrino emissivities.  Here, as a
first application, we will concentrate on the heat capacity.  More
details about the evaluation of the heat capacity and a discussion of
the usually considered contributions can be found in
\cite{Gnedin01}. In what follows, we will concentrate on the
particularly interesting case of the neutron star inner crust.

The core of neutron stars is composed most probably of homogeneous
neutron rich matter, whereas the crust contains different
inhomogeneous structures. The inner crust is thereby characterized by
the transition from a lattice of atomic nuclei in the outer crust to
homogeneous matter in the core. Ravenhall et al.~\cite{Ravenhall83}
and Hashimoto et al.~\cite{Hashimoto84} predicted that this transition
passes via more and more deformed nuclei. Starting from an almost
spherical shape, they could form rods or slabs immersed in a neutron
gas at the different densities. These ``spaghetti'' and ``lasagna''
phases are commonly called the nuclear ``pasta''. At higher densities,
even closer to the core, other phases such as neutron-gas bubbles
inside the dense matter (``swiss cheese'' phase) etc. are
expected. The formation of the different structures strongly depends
on the relative strength of the nuclear surface energy, Coulomb energy
and bulk energy, such that it depends on the nuclear interaction. This
prediction has been confirmed within different models for the nuclear
interaction, see,
e.g.,~\cite{Oyamatsu93,Pethick95,Watanabe03,Avancini09}. These
evaluations have been performed at zero temperature. It is clear that
at some critical temperature the pasta structures will disappear due
to thermal excitations. However, this melting temperature is of the
order of several MeV (see e.g.~\cite{Watanabe00,Avancini10}).

Another point is that in neutron stars older than several minutes,
matter becomes superfluid. A first evidence for superfluidity in
neutron stars has been discussed already in 1969~\cite{Baym69},
shortly after the discovery of the first pulsars, in connection with
the observation of ``glitches''.  Since then much effort has been
devoted to the question of superfluidity and superconductivity in
neutron star matter, for the inner crust as well as for the
homogeneous core, see for example~\cite{Pethick95}.  There is no
consensus on the exact value of the energy gaps $\Delta$ in the inner
crust~\cite{Lombardo00,Gezerlis10}, but the common agreement is that
they are of the order of 1 MeV~\cite{Chamel08}. A pairing gap much
larger than the temperature strongly suppresses the contribution of
individual neutrons to the specific heat which is thus very much
dependent on the pairing strength~\cite{Gnedin01,Fortin09}. For
moderate and strong pairing, the main contributions to the heat
capacity considered so far in the crust are thus electrons and lattice
vibrations as well as collective excitations of nuclei. However, the
superfluid character of neutron star matter induces collective
excitations of the neutron gas, not considered before, which can give
an important contribution to the heat capacity in certain regions, see
~\cite{Aguilera08,Pethick10,Cirigliano11}.

The aim of the present paper is to study these collective excitations
in the inner crust employing a superfluid hydrodynamics approach.
Naturally, there is a vast literature on hydrodynamics for neutron
stars in different contexts, including the effects of superfluidity
\cite{Epstein88,Carter98,Andersson02,Carter05,Gusakov05,Carter06,Carter06b}.
Most of these models are dedicated to the study of macroscopic neutron
star properties, whereas our main aim is to study the excitation
spectrum of the crust on much smaller length scales. In spirit this is
similar to Refs.~\cite{Pethick10,Cirigliano11,Sedrakian96}, where hydrodynamic
equations are developed to study the superfluid Goldstone boson and
(lattice) phonons in the long wavelength limit. However, we are
interested in shorter wavelengths at which effects of the
inhomogeneous structure will manifest themselves, too. In this sense,
our approach is situated in between the long wavelength limit and the
completely microscopic calculations (see
e.g.~\cite{Sandulescu04,Khan05,Fortin09}) employing the Wigner-Seitz
approximation~\cite{Negele71}. In the latter case, the wavelengths are
limited to the size of the Wigner-Seitz cell, because of the imposed
boundary conditions which do not include the coupling between
neighboring cells.

The paper is organized as follows. In Sec.~\ref{sec:formalism}, we
describe the superfluid hydrodynamics approach. We summarize the
hydrodynamic equations, discuss the boundary conditions and the
microscopic input. In Sec.~\ref{sec:results}, we show our first
results. For simplicity, we restrict ourselves to one-dimensional
inhomogeneities (lasagna phase) in this exploratory study. We discuss
the spectrum of the collective modes and their contributions to the
specific heat. A summary and perspectives of our work are exposed in
Section~\ref{sec:summary}.

Throughout the article, $c$, $\hbar$, and $k_B$ denote the speed of
light, the reduced Planck constant, and the Boltzmann constant,
respectively.

\section{Model}
\label{sec:formalism}
\subsection{Superfluid hydrodynamics approach}
In this paper, we are interested in temperatures below $\sim 10^9$ K,
which is very small compared to the gap energy $\Delta$. Therefore we
can use the zero temperature approximation, thus assuming that there
are no normal fluids but only superfluids. In this limit, the dynamics
of a superfluid system with slow temporal and spatial variations is
completely determined by the dynamics of the phase of the order
parameter: If the superfluid order parameter is written as
$\Delta(\vek{r}) = |\Delta(\vek{r})| e^{i\phi(\vek{r})}$, the
superfluid velocity is given by $\vek{v}_s =
(\hbar/2m)\vek{\nabla}\phi$ \cite{LandauLifshitz9}, $m$ being the
nucleon mass. Actually, as pointed out in Ref.~\cite{Chamel06}, the
phase of the order parameter determines the momentum per particle
$\vek{p} = m\vek{v}_s$ and not the fluid velocity $\vek{v}$. This
distinction is important in the context of ``entrainment'' in a system
containing protons and neutrons, see below.

An important length scale is the superfluid coherence length, $\xi_0 =
\hbar v_F/\pi\Delta$ \cite{FetterWalecka}, where $v_F$ denotes
the Fermi velocity. It varies from several fm up to tens of fm for
typical values of the densities, neutron fractions, and gaps in the
inner crust. As can be shown by deriving the equations of superfluid
hydrodynamics from the microscopic time-dependent Bogoliubov-de Gennes
(or Hartree-Fock-Bogoliubov) equations
\cite{ToniniWerner,UrbanSchuck}, the hydrodynamic approach is valid if
the length scale of spatial variations is larger than $\xi_0$, and
frequencies are small compared with $\Delta/\hbar$. As will be discussed
later on, for the concrete examples we consider, we are at the limits
of validity of the approach. However, considering the tremendous
difficulties to perform completely microscopic calculations beyond the
Wigner-Seitz approximation, we leave such investigations for the
future and consider our approach sufficient for the moment.

In addition, we will neglect the Coulomb interaction of the
protons. This represents an enormous simplification, but at the same
time it implies that we cannot correctly reproduce the phonons of the
Coulomb lattice. In homogeneous matter, too, the Coulomb interaction
plays an important role for the collective modes, in particular the
coupling of the proton plasmon mode with the electrons as discussed in
\cite{Baldo08,Baldo09}, but it is beyond the scope of the present
paper. Our main focus lies therefore on the dynamics of the neutron
gas, which is however coupled to the proton dynamics due to the
nuclear interaction.

In principle, the equations of superfluid hydrodynamics can be derived
from the underlying microscopic theory, as it was done for the case of
ultracold trapped fermionic atoms in
\cite{ToniniWerner,UrbanSchuck}. Here, we follow the simpler way to
derive them from local conservation laws. Since the fluid velocities
and the densities are low enough, we will use a non-relativistic
formulation.

The first conservation law is neutron and proton number
conservation \footnote{On the time scales of the collective
  oscillations we want to study, weak interaction processes
  transforming neutrons and protons into each other can be safely
  neglected.}. This results in two continuity equations, one for
neutrons ($a = n$) and one for protons ($a = p$),
\begin{equation}
\partial_t n_a + \vek{\nabla}\cdot(n_a\vek{v}_a) = 0\,, \\
\label{eq:continuite}
\end{equation}
where $n_a$ denotes the particle number density of species $a$.

The second conservation law, the conservation of momentum, results in
the Euler equations, which can be written as
\begin{equation}
n_a \left(\partial_t \vek{p}_a + \vek{\nabla} \tilde{\mu}_a \right) =
0\, ,
\label{eq:euler}
\end{equation}
where $\tilde{\mu}_a$ is the rest-frame chemical potential defined as
the conjugate momentum with respect to the particle density $n_a$ in a
variational approach~\cite{Prix04}. The explicit expression is
\begin{equation}
\tilde{\mu}_a = \mu_a + \vek{v}_a\cdot \vek{p}_a - \frac{1}{2} m_a
v_a^2\,,
\label{eq:murestframe}
\end{equation}
where $\mu_a$ is the local chemical potential of species $a$. Due to
the interaction between neutrons and protons, $\mu_a$ depends on the
densities of both species.

In pure neutron matter, the momentum $\vek{p}_n$ is simply given by
$\vek{p}_n = m_n\vek{v}_n$. However, in a system containing neutrons
and protons, the two species drag each other due to their
interaction. In the theory of superfluids, this effect is known as
entrainment~\cite{Andreev75}. As a consequence, fluid momenta are
misaligned with particle velocities. The relationship between the
velocity and the momentum can be expressed via the entrainment matrix
(also called Andreev-Bashkin or mass-density matrix) \cite{Chamel06}:
\begin{equation}
m_a n_a \vek{v}_a = \sum_{b = n,p} \rho_{ab} \frac{\vek{p}_b}{m_b}\,.
\label{eq:entrainement}
\end{equation}
In practice, at densities which are relevant in the inner crust, the
non-diagonal elements of $\rho$ are small \cite{Chamel06}, i.e.,
\begin{equation}
\rho_{ab} \approx m_a n_a \delta_{ab}\,.
\end{equation}

\subsection{Microscopic input}
\label{sec:microscopicinput}
As microscopic input, we need the equation of state, i.e., the
relation between the densities $n_a$ and the chemical potentials
$\mu_a$, and the entrainment matrix $\rho$. In our concrete numerical
examples, we will use the results of the work by Avancini et
al.~\cite{Avancini09} for the equilibrium configurations. They
evaluate the structure of the pasta phases for charge neutral matter
in $\beta$ equilibrium using a density dependent relativistic
mean-field model, the DDH$\delta$ model (originally called
DDH$\rho\delta$)~\cite{Gaitanos04,Avancini04,Avancini09}, for the
nuclear interaction. In order to be consistent, we shall calculate the
chemical potentials $\mu_a$ and the entrainment matrix $\rho$ with the
same interaction. For the entrainment matrix, we closely follow
Gusakov et al.~\cite{Gusakov09}, who generalized the determination of
the entrainment matrix for neutron-proton mixtures based on
Landau-Fermi liquid theory~\cite{Borumand96} to relativistic
models. The only modification of the expressions in
Ref.~\cite{Gusakov09} we have to perform is due to the presence of the
isovector-scalar $\delta$ meson in the DDH$\delta$ model, which
modifies the Dirac effective nucleon mass. In particular, the latter
is no longer the same for neutrons and protons. Since our hydrodynamic
equations are formulated non-relativistically, we consider only the
non-relativistic limit of the entrainment matrix ($\rho_{ab} = m_a m_b
c^2 Y_{ab}$ in the notation of \cite{Gusakov09}).
\subsection{Linearization around stationary equilibrium}
\label{sec:linearisation}
In order to proceed we will linearize Eqs.~(\ref{eq:continuite}) and
(\ref{eq:euler}) around stationary equilibrium. Let us write the
different quantities as a sum of their equilibrium value and a
perturbation, $X = X_\eq + \delta X$ (in the case of the velocities
and momenta we will write the perturbation simply as $\vek{v}_a$ and
$\vek{p}_a$, respectively, since the equilibrium values of these
quantities are zero). The equations can be simplified a lot, since all
temporal and spatial derivatives of equilibrium quantities vanish
(except at phase boundaries, which will be treated in the next
subsection). Eqs.~(\ref{eq:continuite}) and (\ref{eq:entrainement}) then
reduce to
\begin{equation}
\partial_t \delta n_a = - \sum_{b=n,p} \frac{\rho_{ab,\eq}}{m_a m_b}
\vek{\nabla} \cdot \vek{p}_b\,,
\label{eq:delta_continuite}
\end{equation}
and Eqs.~(\ref{eq:euler}) and (\ref{eq:murestframe}) become
 \begin{equation}
\partial_t \vek{p}_a = - \vek{\nabla} \delta \mu_a\,.
\label{eq:delta_euler}
\end{equation}
We will now express the variation of the densities in
Eq.~(\ref{eq:delta_continuite}) in terms of the variation of the
chemical potentials, 
\begin{equation}
\delta n_a = \sum_{b = n,p} J_{ab} \delta \mu_b\,,
\end{equation}
where
\begin{equation}
J_{ab} = \Big(\frac{\partial n_a}{\partial \mu_b}\Big)_\eq\,.
\end{equation}
Inserting the resulting equation into the divergence of
Eq.~(\ref{eq:delta_euler}) one obtains the following system of two
coupled wave equations for $\delta\mu_n$ and $\delta\mu_p$:
\begin{equation}
\sum_{b=n,p}(KJ)_{ab}\,\partial_t^2 \delta \mu_b = \nabla^2 \delta \mu_a\,,
\label{eq:sound_wave}
\end{equation}
where $K$ is the inverse of the matrix
\begin{equation}
(K^{-1})_{ab} = \frac{\rho_{ab,\eq}}{m_a m_b}\,.
\end{equation}
The coupling
arises from the non-diagonal elements of the matrices $J$ and $K$ due
to the neutron-proton interaction. Let us now make the ansatz that the
perturbations have the form of a plane wave, $\delta\mu_a(\vek{r},t) =
U_a e^{-i\omega t + i\vek{k}\cdot \vek{r}}$. Eq.~(\ref{eq:sound_wave})
can then be written as a $2\times 2$ eigenvalue problem
\begin{equation}
 \sum_{b = n,p} (K J)_{ab} U_b = \frac{1}{u^2} U_a\,,
\label{eq:eigenvalue}
\end{equation}
with $u = \omega/k$ denoting the sound velocity. The two eigenvalues
give two sound velocities which we will label $u^\pm$. Note that the
corresponding eigenvectors, $U_a^\pm$, do not describe pure proton or
neutron waves, but combinations of both. We denote by $+$ and $-$ the
modes where neutrons and protons oscillate in phase and out of phase,
respectively.

In the special case of pure neutron matter, there is only one mode,
which can be obtained from the above equations by setting $n_p = 0$.
Its sound velocity is given by
\begin{equation}
u^2 = \Big(\frac{n_n}{m_n}\frac{\partial\mu_n}{\partial n_n}\Big)_\eq\,.
\label{uneutron}
\end{equation}
\subsection{Boundary conditions}
\label{sec:boundary}
In our model, we consider the inhomogeneous phases in the inner crust
as mixed phases where a neutron gas (phase 1) coexists with a dense
phase (phase 2) containing neutrons and protons. However, in order not
to have to write everything separately for phase 1 and phase 2, we
will write all equations, unless otherwise stated, for the general
case that neutrons and protons are present in both phases. The
equations relevant for phase 1 can easily be obtained by considering
the special case $n_{p1} = 0$. The fact that both phases coexist
implies that in equilibrium the chemical potentials and pressures are
equal in both phases: $\mu_{a1} = \mu_{a2}$ and $P_1 = P_2$. The
description of the interface between the two phases requires a
microscopic formalism and is beyond the scope of this work.

In our model, we assume that the hydrodynamic equations are valid in
both the gas and the dense phase, but since they do not say anything
about the behavior at the interface, they have to be supplemented by
appropriate boundary conditions. The first boundary condition arises
from the obvious requirement that contact has to be maintained at all
times at the interface~\cite{wave_propagation}. Therefore, the
displacement normal to the interface has to be continuous and equal
for all components ($a=n,p$) at all times. Hence, the velocities
normal to the interface must satisfy:
\begin{equation}
v_{\perp n1}(\vek{r}) = v_{\perp p1}(\vek{r}) = 
v_{\perp n2}(\vek{r}) = v_{\perp p2}(\vek{r})\,.
\label{boundarycondition1}
\end{equation}

The second boundary condition arises from the requirement that the
pressure $P$ on both sides of the interface must be equal \cite{LaHeurte83}:
\begin{equation}
P_1(\vek{r}) = P_2(\vek{r})\,.
\end{equation}
If we linearize this condition, it can be written as
\begin{equation}
\sum_{a=n,p} n_{a1}(\vek{r}) \delta\mu_{a1}(\vek{r})
  = \sum_{a=n,p} n_{a2}(\vek{r}) \delta\mu_{a2}(\vek{r})\,,
\label{boundarycondition2}
\end{equation}
where the index $\eq$ after $n_{a1}$ and $n_{a2}$ has been dropped for
brevity.

Before applying our model to the neutron-star inner crust, let us see
whether these boundary conditions give reasonable results for
collective modes in isolated nuclei. For simplicity, we will consider
a nucleus with equal numbers of neutrons and protons ($N = Z =
A/2$). Within the hydrodynamic model, the nucleus is a homogeneous
sphere with a sharp surface at $r = R$. The proton and neutron
densities inside the nucleus are $n_n = n_p = n_0/2$, where $n_0 =
0.153$ fm$^{-3}$ is the saturation density within the DDH$\delta$
model. As a first example, we consider the isoscalar monopole mode,
where neutrons and protons oscillate together in radial direction. The
solution of the wave equation inside the nucleus is $\delta\mu_n =
\delta\mu_p \propto j_0(\omega r/u^+)$, where $j_l$ is a spherical
Bessel function and $u^+ = 0.169\,c$ is the sound velocity for the
in-phase oscillation of neutrons and protons. Since protons and
neutrons move together, the first boundary condition
(\ref{boundarycondition1}) is automatically satisfied, while the
second one, Eq.~(\ref{boundarycondition2}), requires $\delta\mu(r=R) =
0$. Consequently, the energy of the monopole mode is $\hbar\omega =
\pi \hbar u^+/R \approx 90$ MeV$/A^{1/3}$.

Another interesting simple case is the isovector giant-dipole
resonance (GDR), where neutrons and protons oscillate against each
other in $z$ direction. In this case, our approach is identical to the
Steinwedel-Jensen model of the GDR \cite{SteinwedelJensen}. Again, the
solution of the wave equation is straight-forward and gives
$\delta\mu_n = -\delta\mu_p \propto j_1(\omega r/u^-) \cos\theta$,
where $\theta$ is the angle between $\vek{r}$ and the $z$ axis, and
$u^- = 0.233\,c$ is the sound velocity for the out-of-phase
oscillation of neutrons and protons. In this case, the second
condition (\ref{boundarycondition2}) is automatically fulfilled, but
now the first one, Eq.~(\ref{boundarycondition1}), becomes
relevant. Using the Euler equation (\ref{eq:delta_euler}), one can
show that the radial component of the velocity field is proportional
to $v_{r n} = -v_{r p}\propto \partial\delta\mu/\partial r\propto
j_1^\prime(\omega r/u^-) \cos \theta$, so that Eq.~(\ref{boundarycondition1})
gives $\hbar\omega = 2.08 \hbar u^-/R \approx 82$ MeV$/A^{1/3}$.

These results for the isoscalar monopole and the isovector GDR are
quite reasonable, at least for heavy nuclei, although their energies
are much higher than the pairing gap $\Delta$, such that superfluid
hydrodynamics should strictly speaking not be applicable. The reason
is that for these particular resonances (contrary to, e.g., the
quadrupole mode \cite{RingSchuck}) the Fermi-surface distortion does
not play any role, so that hydrodynamics works even in the normal
phase without pairing. We conclude that, at least in some cases, the
limits of validity of the hydrodynamic approach may be interpreted
very generously.

\subsection{Collective modes in a periodic slab structure}
\begin{figure}
\begin{center} 
\includegraphics[width=2cm,angle=-90]{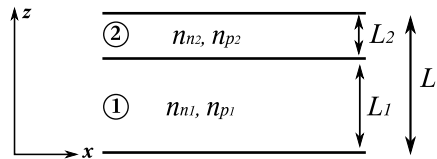}
\caption{Diagram representing the slab structure.}
\label{fig:lasagne}
\end{center}
\end{figure}

Because of their electric charge, the droplets (or rods, or slabs) of
the dense phase arrange in a regular periodic lattice in order to
minimize the Coulomb energy. Charge neutrality on a macroscopic scale
is guaranteed by the presence of an almost uniform, strongly
degenerate electron gas. The size and form of the structures is
determined by the interplay of Coulomb energy (favoring small
structures) and surface energy (favoring large structures). Since both
the Coulomb and the surface energy are neglected in our approach, the
determination of the size and form of the structures in equilibrium is
beyond the scope of our work. Instead, we will consider the
equilibrium geometry as input and calculate the collective
oscillations in this geometry. For simplicity, we will restrict
ourselves to the simplest geometry which is a structure of
periodically alternating slabs with different proton and neutron
densities as illustrated in Fig.~\ref{fig:lasagne} (lasagna phase). To
be specific, we will consider the slabs to be perpendicular to the $z$
axis.

Our aim is to describe the collective modes of this structure. The
equilibrium properties of the structure itself, i.e., the densities
$n_{n1}$, $n_{p1}$, $n_{n2}$ and $n_{p2}$, and the slab thicknesses
$L_1$ and $L_2$, are input parameters which we take from
Ref.~\cite{Avancini09}. The excitations are then obtained by solving
in each slab the wave equation (\ref{eq:sound_wave}) together with the
boundary conditions (\ref{boundarycondition1}) and
(\ref{boundarycondition2}) at the interfaces between neighboring
slabs.

At each phase boundary, the waves will be partially (or totally)
reflected. It is therefore not sufficient to make a plane-wave ansatz
in each slab, but one has to consider the reflected wave, too. Thus,
we can make the following ansatz in each slab:
\begin{equation}
\delta \mu_a(\vek{r},t) = \sum_{\sigma=\pm} 
  e^{-i\omega t+i\vek{k}_\paral\cdot \vek{r}_\paral}
  \left( \alpha^\sigma e^{ik_z^\sigma z} + \beta^\sigma e^{-i k_z^\sigma z}\right)
  U_a^\sigma~,
\label{eq:superposition}
\end{equation}
where $U_a^\pm$ denote the normalized eigenvectors of
Eq.~(\ref{eq:eigenvalue}), $\vek{k}_\paral = (k_x,k_y,0)$ and
$\vek{r}_\paral = (x,y,0)$ are the components of $\vek{k}$ and $\vek{r}$
parallel to the slab, and the $k_z^\pm$ have to satisfy
\begin{equation}
k_z^{\pm\, 2} = \frac{\omega^2}{u^{\pm\,2}} - k_\paral^2\,.  
\end{equation}
Note that $k_z^\pm$ can be real or imaginary. The velocities
$\vek{v}_a$ can be expressed in terms of the coefficients $\alpha^\pm$
and $\beta^\pm$, too. By using the Euler equation (\ref{eq:delta_euler}),
one finds that for a plane wave with wave vector $\vek{k}$ the
velocity field is given by
\begin{equation}
\vek{v}_a = \frac{\vek{k}}{n_a\omega} \sum_b (K^{-1})_{ab}\delta\mu_b\,.
\end{equation}
If we define
\begin{equation}
V^\pm_a = \frac{1}{n_a}\sum_b (K^{-1})_{ab} U^\pm_b\,,
\end{equation}
the superposition of plane waves according to
Eq. (\ref{eq:superposition}) gives
\begin{equation}
v_{za} = \sum_{\sigma=\pm} \frac{k_z^\sigma}{\omega} 
  e^{-i\omega t+i\vek{k}_\paral\cdot\vek{r}_\paral} 
  \left( \alpha^\sigma e^{ik_z^\sigma z} 
    - \beta^\sigma e^{-i k_z^\sigma z}\right) V_a^\sigma
\end{equation}
and a similar relation for $\vek{v}_\paral$.

The next step is to determine the coefficients $\alpha^\pm$ and
$\beta^{\pm}$ by matching the solutions in neighboring slabs according
to the boundary conditions. If we use indices 1, 2, 3 in order to
indicate the quantities in three consecutive slabs, we have
perturbations $\delta \mu_{a1}$ valid for $0 < z< L_1$, $\delta
\mu_{a2}$ for $L_1<z<L \equiv L_1+L_2$, and $\delta \mu_{a3}$ for $L <
z < L+L_1$ with four unknown amplitudes, $\alpha^\pm$ and $\beta^\pm$,
in each slab. Note that due to the periodicity the equilibrium
properties of slab 3 are equal to those of slab 1, but the
coefficients $\alpha^\pm$ and $\beta^\pm$ are in general different in
slabs 1 and 3.

Written explicitly, the boundary conditions are
\begin{gather}
v_{zp2}(z=L_1) = v_{zn2}(z=L_1)\,,\nonumber\\
v_{zn1}(z=L_1) = v_{zn2}(z=L_1)\,,\nonumber\\
v_{zp1}(z=L_1) = v_{zn2}(z=L_1)\,,\nonumber\\
\sum_{a= n,p} n_{a1}\delta\mu_{a1}(z=L_1) =
  \sum_{a= n,p} n_{a2}\delta\mu_{a2}(z=L_1)\,,
\label{condition1}
\end{gather}
and four analogous equations relating quantities of slabs 2 and 3 at
$z = L$.

It is evident that, in order to satisfy the conditions for all
$\vek{r}_\paral$, the components $\vek{k}_\paral$ must be equal in all
three slabs, i.e.,
\begin{equation}
\vek{k}_{\paral 1} = \vek{k}_{\paral 2} = \vek{k}_{\paral 3} 
  \equiv \vek{q}_\paral\,.
\label{eq:kqparal}
\end{equation}

So far, the boundary conditions give us eight equations for the twelve
unknown coefficients $\alpha^\pm_1, \dots, \beta^\pm_3$. In order to
close the system of equations, we have to take into account
translational invariance of the system. This can be expressed via the
Floquet-Bloch theorem~\cite{Floquet}:
\begin{equation}
\delta \mu_a(\vek{r} + \vek{R},t) = e^{i \vek{q} \cdot \vek{R}}\,
\delta \mu_a(\vek{r},t)\,,
\label{eq:bloch}
\end{equation}
where $\vek{q}$ is the Bloch momentum and $\vek{R} = (R_x,R_y,R_z)$ is
a vector such that the system is invariant under a shift
$\vek{r}\to\vek{r}+\vek{R}$. In our case, $\vek{R}_\paral$ can be
arbitrary, but $R_z$ has to be a multiple of the periodicity $L$, see
Fig.~\ref{fig:lasagne}. With respect to $\vek{R}_\paral$, the
condition (\ref{eq:bloch}) is automatically satisfied due to
Eq.~(\ref{eq:kqparal}). But in the case $\vek{R} = (0,0,L)$,
Eq.~(\ref{eq:bloch}) implies in particular
\begin{equation}
\delta\mu_{a3}(x,y,z=L,t) = e^{iq_z L}\,\delta\mu_{a1}(x,y,z=0,t)
\end{equation}
and an analogous relation for the velocity. Inserting this into the
boundary conditions at the interface between slabs 2 and 3 at $z = L$,
we obtain
\begin{gather}
v_{zp2}(z=L) = v_{zn2}(z=L)\,,\nonumber\\
e^{iq_z L}\,v_{zn1}(z=0) = v_{zn2}(z=L)\,,\nonumber\\
e^{iq_z L}\,v_{zp1}(z=0) = v_{zn2}(z=L)\,,\nonumber\\
e^{iq_z L}\sum_{a=n,p} n_{a1}\delta\mu_{a1}(z=0) =
  \sum_{a=n,p} n_{a2}\delta\mu_{a2}(z=L)\,,
\label{condition3}
\end{gather}
i.e., we have now a system of eight equations, Eqs. (\ref{condition1})
and (\ref{condition3}), for eight coefficients
$\alpha^\pm_1,\dots,\beta^\pm_2$.

This system of equations has a non-trivial solution if the determinant
of the corresponding $8\times 8$ matrix vanishes. For a given choice
of $q_\paral$ and $q_z$ ($q_z$ may be limited to the first Brillouin
zone, i.e., $-\pi/L < q_z < \pi/L$), this gives us an equation for
$\omega$ with an infinite number of discrete solutions.

Note that, as mentioned before, in the case we will actually consider,
the proton density vanishes in slab 1 (and 3). In this case, the
proton velocity is not defined in that slab and we have only two
coefficients $\alpha_1$ and $\beta_1$ instead of four coefficients
$\alpha^\pm_1$ and $\beta^\pm_1$, since in pure neutron matter there
is only one eigenmode instead of two. The number of equations is also
reduced by two, since the third equation of Eqs.~(\ref{condition1})
and the third equation of Eqs.~(\ref{condition3}) can be removed. We
are therefore left with a $6\times 6$ instead of $8\times 8$
problem. In this case, it is interesting to notice that there are two
different types of modes: Modes propagating through all slabs, whose
energies depend on $q_z$, and modes of the dense slabs (slab~2) only,
whose energies are independent of $q_z$. The latter are modes
where protons and neutrons oscillate against each other such that at
the boundaries $(z = L_1$ and $z = L$) $v_{zn}$, $v_{zp}$, and $\delta
P$ vanish simultaneously (analogous to the isovector GDR in an
isolated nucleus, discussed at the end of the previous subsection).

When looking for the solutions for $\omega$, one has to be careful to
retain only physical solutions. It is easy to see that if one of the
three wave numbers $k_{z1}$, $k^+_{z2}$, or $k^-_{z2}$ vanishes, i.e.,
if $\omega/q_\paral$ equals one of the three sound velocities $u_1$,
$u^+_2$, or $u^-_2$, the system of equations is solved by choosing the
corresponding coefficients as $\alpha = -\beta$ and setting all the
other coefficients equal to zero. However, this solution implies
$\delta\mu = 0$ and therefore does not correspond to a physical
excitation.

\section{Results for the lasagna phase}
\label{sec:results}
\subsection{Excitation spectrum}
\label{sec:excitations}
Let us now investigate the resulting excitation spectrum for a specific
example. As mentioned before, the values for the equilibrium quantities will
be taken from the work by Avancini et al.~\cite{Avancini09}, who have studied
the structure of pasta phases in a relativistic mean field model.  Our
geometry corresponds to the lasagna phase, appearing close to the transition
to uniform matter in the core, which has been found in Ref.~\cite{Avancini09}
in the case of zero temperature and $\beta$-equilibrium for baryon number
densities $ 0.077$ fm$^{-3} \lesssim n_B \lesssim 0.084$ fm$^{-3}$, in good
agreement with the results by Oyamatsu~\cite{Oyamatsu93}. For our example we
have chosen an intermediate density, $n_B = 0.08$ fm$^{-3}$. The corresponding
properties of the two phases 1 and 2 are listed in Table~\ref{tab:structure}.
\begin{table}[t]
\caption{\label{tab:structure} Properties of the lasagna phase within
  the model by Avancini et al. \cite{Avancini09} studied in our
  example. The average densities of the total system are given by $n_n
  = (L_1/L) n_{n1}+(L_2/L) n_{n2}$ etc. Baryon density and proton
  fraction are defined as $n_B = n_n+n_p$ and $Y_p = n_p/n_n$,
  respectively.}
\begin{ruledtabular}
\begin{tabular}{ccccc}
      &                     & slab 1 & slab 2 & total \\
\hline
$L$             & (fm)      & 9.40          & 7.38         & 16.78 \\
$n_n$           & (fm$^{-3}$) & 0.0701       & 0.0885       & 0.0782 \\
$n_p$           & (fm$^{-3}$) & 0            & 0.0041       & 0.0018 \\
$n_B = n_n+n_p$ &(fm$^{-3}$) & 0.0701        & 0.0926       & 0.0800 \\
$Y_p = n_p/n_B$ &            & 0            & 0.0447       & 0.0227 \\
$u$ or $u^+$    & ($c$)     & 0.0641        & 0.0354       &        \\
$u^-$           & ($c$)     &               & 0.1369       &        
\end{tabular}
\end{ruledtabular}
\end{table}

With the actual numbers for the densities and the dimensions of the
structure, the coherence length for a gap of 1 MeV is of the same
order of magnitude as the size of the layers, i.e. the scale for
spatial variations. That means that our superfluid hydrodynamics
approach touches its limit of validity for this example. Strictly
speaking, we should also limit ourselves to energies which are small
compared to $\Delta$. However, there are many cases where the
hydrodynamic approach works reasonably well although its initial
assumptions are not fulfilled. Examples are the dipole and monopole
resonances in ordinary nuclei mentioned in the preceding section, or
the ``supergiant resonances'' in spherical Wigner-Seitz cells used to
model the neutron-star inner crust, whose excitation energies agree
well with an estimate obtained from the sound velocity of the
hydrodynamic Bogoliubov-Anderson mode \cite{Khan05}.

\begin{figure*}
\begin{center} 
\includegraphics[width=12cm]{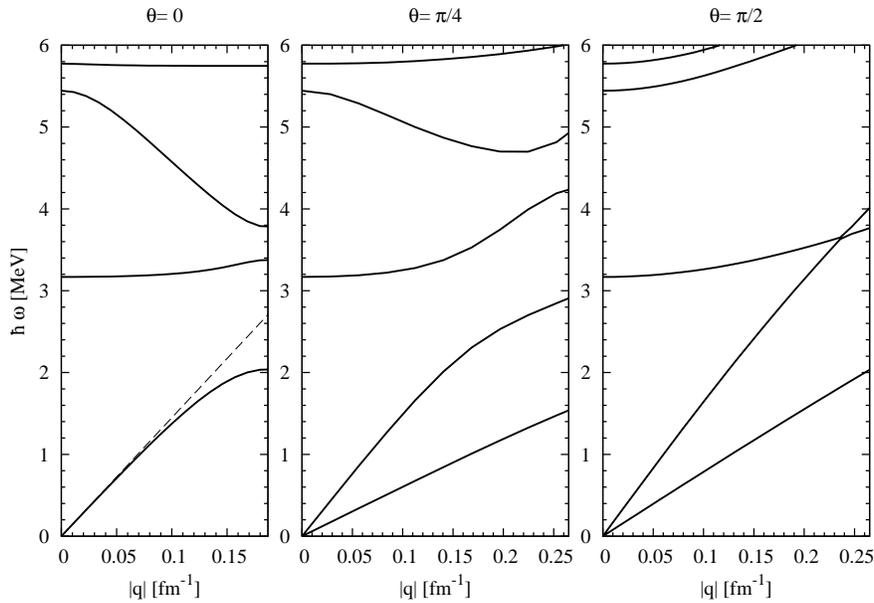}
\caption{Dispersion relation of the modes propagating along the
  $z$-axis ($\theta = 0$, left), at an angle of $45^\circ$ ($\theta =
  \pi/4$, center), and in the plane parallel to the slabs ($\theta =
  \pi/2$, right). The dashed line in the left panel
  corresponds to the approximation Eq.~(\ref{acoustic3d}).}
\label{fig:dispersion}
\end{center}
\end{figure*}

After this remark of caution, let us discuss the solutions for the
energies $\omega$ shown in Fig.~\ref{fig:dispersion}
as functions of $q \equiv |{\bm q}|$ for three different angles $\theta$ between
$\vek{q}$ and the $z$ axis (i.e., $q_z = q \cos\theta$ and $q_\paral =
q\sin\theta$). The left panel shows the dispersion relation for waves
propagating in $z$-direction, i.e. perpendicular to the interfaces
between the different slabs. One observes an acoustic branch with an
approximately linear dispersion law
\begin{equation}
\omega = u_s q \label{acoustic3d}
\end{equation}
at low energies, and several optical branches with a finite energy for
$q = 0$, analogously to phonons branches in a crystal.

Note that within the Wigner-Seitz approximation, which is usually
employed in microscopic calculations
\cite{Khan05,Sandulescu04,Fortin09}, we would only obtain a discrete spectrum
corresponding to our spectrum in the case $q = 0$. The reason is that 
in this approximation the coupling between
cells is neglected, and thus each cell has the same excitation spectrum. The
degeneracy of the modes in each cell is lifted by the
coupling between cells, which gives rise to a momentum dependent spectrum as
obtained in our approach.

The slope of the acoustic branch, i.e., the speed of sound, coincides
(see dashed line in Fig.~\ref{fig:dispersion}) with the usual
thermodynamic expression for the sound velocity
\begin{equation}
u_s^2 = \frac{1}{m}\, \frac{\partial P}{\partial n_B}\Big|_{Y_p}\,,
\end{equation}
where $n_B$ is the average baryon density of the inhomogeneous phase.
To evaluate this derivative, we squeeze or expand our unit cell of
length $L$ by a small amount $\delta L = \delta L_1+\delta L_2$. From
the requirement $\delta P_1 = \delta P_2 = \delta P$ we can determine
$\delta L_1$ and $\delta L_2$ and thus $\delta P$. The final result
can be written in a compact form as
\begin{equation}
\frac{L}{n_B u_s^2} = \frac{L_1}{n_{B1} u_{s1}^2}
  + \frac{L_2}{n_{B2} u_{s2}^2}\,,
\end{equation}
where we have defined for each phase $i = 1,2$ 
\begin{equation}
u_{si}^2 = \frac{1}{m}\,\frac{\partial P_i}{\partial n_{Bi}}\Big|_{Y_{pi}}\,.
\end{equation}
Note that $u_{s1}$ is identical to the sound velocity $u_1$
[cf. Eq.~(\ref{uneutron})], whereas $u_{s2}$ is different from the two
sound velocities $u^\pm_2$.

This linear branch corresponds roughly to the long wavelength limit
discussed in Ref.~\cite{Pethick10, Cirigliano11} although, of course,
the numerical value of the sound speed is not the same because we
neglect elastic effects of the proton lattice due to Coulomb
interaction. At higher wave vectors $q$, there are deviations from the
linear behavior related to the inhomogeneous structure. At these
energies the long wavelength limit is no longer valid.

The central and right panels of Fig.~\ref{fig:dispersion} show the
excitation spectrum for different values of the angle, $\theta =
\pi/4$ and $\pi/2$, respectively. One observes that the slope of the
acoustic branch discussed before changes: in the present example,
$u_s$ increases from $0.072\,c$ in the case $\theta = 0$ to $0.085\,c$
in the case $\theta = \pi/2$. The reason is that the wave, which is
perfectly longitudinal ($\vek{v} \parallel \vek{q}$) in the case $\theta =
0$, becomes more complicated in the case $\theta \neq 0$ and the
nucleons oscillate now in both longitudinal and transverse
directions. But the most important consequence of non-zero angle
$\theta$ is the appearance of a second acoustic branch, whose slope is
strongly angle dependent. In fact, if one writes the energy of this
new branch as
\begin{equation}
\omega = u_s^\prime q_\paral = u_s^\prime q \sin\theta\,, \label{acoustic2d}
\end{equation}
the ``two-dimensional sound velocity'' $u_s^\prime$ defined by this
equation depends only weakly on $q_z$ and $q_\paral$: in the present
example, $u_s^\prime$ varies between $0.04\,c$ for $q_z \ll q_\paral$ and
$0.046\,c$ for $q_\paral \ll q_z$. A detailed analysis of the solutions for
the coefficients $\alpha$ and $\beta$ corresponding to this branch
shows that in this mode, the protons and neutrons oscillate
practically only in the direction parallel to the slabs (i.e., $v_z
\approx 0$), and the motion takes essentially place in the dense
phase.

\subsection{Application to specific heat}
\label{sec:cv}
We are interested here in the contribution of the above discussed
excitation modes to the specific heat. The specific heat, the heat
capacity for constant volume per unit volume, is defined as
\begin{equation}
c_v(T) = \frac{\partial \epsilon}{\partial T}  \Big|_n\,,
\label{eq:cv}
\end{equation}
where $\epsilon$ denotes the energy density. The contribution of the
collective modes can be calculated as follows:
\begin{equation}
\epsilon(T) = \int_{-\pi/L}^{\pi/L} \frac{dq_z}{2\pi}
  \int\frac{d^2q_\paral}{(2\pi)^2} \sum_i \hbar\omega_i(\vek{q})
\frac{1}{e^{\hbar\omega_i(\vek{q})/k_BT}-1}\,.
\end{equation}
Note that we suppose here that the energies $\omega_i(\vek{q})$
depend only very weakly on temperature such that it is justified to
neglect their temperature dependence. This should be a good
approximation as long as the temperature stays well below the value of
the energy gap and we therefore have no significant contribution from
a normal fluid. Another type of temperature dependence could arise
from a change in the structure of the pasta phases. At the
temperatures considered here, however, we do not expect a significant
effect either since the structure starts to be modified considerably
only at higher temperatures~\cite{Watanabe00, Avancini10}.

\begin{figure}
\centering
\includegraphics[width=8cm]{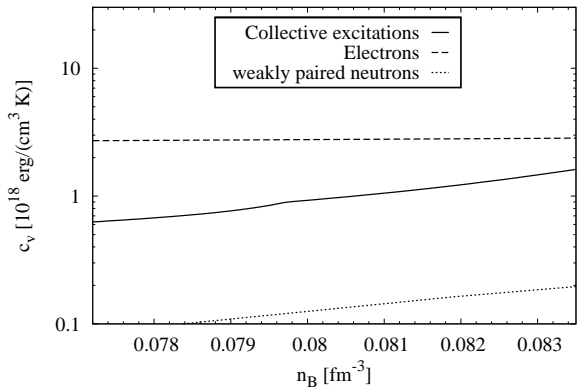}
\caption{Different contributions to the specific heat in the density
  range where one expects to find the lasagna phase, for $T = 10^9$ K. Solid: collective modes, dashed: electrons, dotted: neutron quasiparticles (from Ref.~\cite{Fortin09}).
  Concerning the conversion between astrophysical and nuclear units,
  note that $10^9$ K $ = 86.17 k_B^{-1}$ keV and $10^{18}$ erg K$^{-1}$
  cm$^{-3} = 7.246\times 10^{-6} k_B$ fm$^{-3}$.}
\label{fig:Cv_dens}
\end{figure}

In Fig.~\ref{fig:Cv_dens} 
we show the different contributions to the specific heat in the
density range where the model by Avancini et al.~\cite{Avancini09}
predicts the lasagna phase, for a typical temperature of $10^9$
K. Besides the contribution of the collective modes (solid line), we
display for comparison the contribution of the electrons (dashed line),
which are considered as a practically uniform ultra-relativistic
($\mu_e \gg m_e c^2$) ideal Fermi gas with number density $n_e =
n_p$. At low temperature, the electron gas is strongly degenerate and
its contribution to the specific heat reads
\begin{equation}
c_v^{\mathrm{el.}} = \frac{k_B^2\mu_e^2T}{(\hbar c)^3}\,.\label{Cv_el}
\end{equation}
The importance of the collective modes becomes clear if one considers
the contribution of the gapped neutron quasiparticles (dotted
curve), taken from Ref.~\cite{Fortin09}: In the absence of collective
modes, an excitation of the neutron gas requires the breaking of
Cooper pairs, which is suppressed by a factor of the order of
$e^{-\Delta/k_BT}$. Even in the case of weak pairing, at the present
temperature, this contribution is suppressed by approximately one
order of magnitude with respect to the contribution of the collective
modes.

\begin{figure}
\centering
\includegraphics[width=8cm]{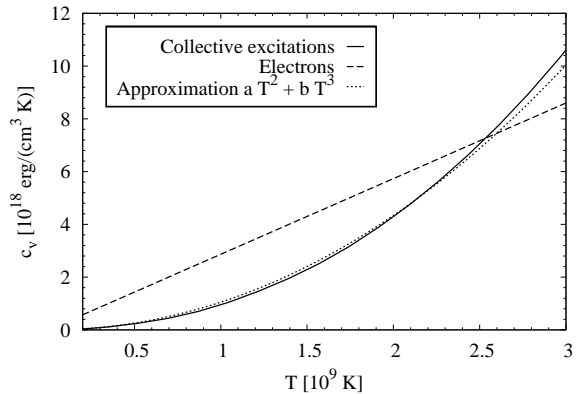}
\caption{Temperature dependence of the contribution of the
  collective modes (solid line) and of the electrons (dashes) to the specific heat for
  the the example studied in the preceding subsection (see
  Table~\ref{tab:structure}). The approximate formula , $a T^2 + b T^3$, see
  Eqs. (\ref{Cv_T3}) and (\ref{Cv_T2}), is shown as a dotted line. }
\label{fig:Cv_T}
\end{figure}

In Fig.~\ref{fig:Cv_T},
we show the temperature dependence of the specific heat corresponding
to the intermediate-density case discussed in
Sec.~\ref{sec:excitations} (solid line). For comparison, we again
display the specific heat due to the electrons (dashed line). Due to
its linear temperature dependence, Eq.~(\ref{Cv_el}), the electron
contribution is always dominant at low temperature, but at higher
temperature, the contribution of the collective modes is comparable or
even larger than the electron contribution.

At the low temperatures considered here, which are well below the
energy of the first optical branch, the contribution of the collective
modes to the specific heat is completely dominated by the two linear
branches discussed in the preceding subsection. As is well known
\cite{Statistical5}, the specific heat due to an acoustic branch with
a linear dispersion relation, Eq.~(\ref{acoustic3d}), reads
\begin{equation}
c_v = \frac{2\pi^2 k_B^4T^3}{15 \hbar^3u_s^3} \equiv b T^3\,.
\label{Cv_T3}
\end{equation}
In the present case, however, we have seen that there is in addition a
``two-dimensional'' branch which propagates only parallel to the slabs
and whose dispersion relation is approximately given by
Eq.~(\ref{acoustic2d}). The contribution of such a mode to the
specific heat is readily shown to be
\begin{equation}
c_v = \frac{3\zeta(3)k_B^3 T^2}{\pi\hbar^2 u_s^{\prime\,2} L} \equiv aT^2\,,
\label{Cv_T2}
\end{equation}
where $\zeta$ is the Riemann zeta function [$\zeta(3) =
1.202\dots$]. Due to its quadratic temperature dependence, this is the
next dominant contribution at low temperatures after the
electrons. 
The result of the simple formula $a T^2+b T^3$, where $a$ and
$b$ have been calculated with the average values $u_s = 0.078\,c$
and $u_s^\prime = 0.042\,c$, is shown as the dotted line in
Fig.~\ref{fig:Cv_T}. Up to the temperatures considered here, it fits
reasonably well the full calculation.

\section{Summary}
\label{sec:summary}
In this paper, we have presented a formalism of superfluid
hydrodynamics to treat density-wave propagation in inhomogeneous
pasta-like nuclear structures which appear in the inner crust of
neutron stars. To account for the periodicity of the structure, we
incorporate the Floquet-Bloch boundary conditions. The idea is
somewhere in between the approaches of Refs.~\cite{Pethick10,
  Cirigliano11}, considering only the long-wavelength limit, averaging
over the microscopic details of the structure, and microscopic
calculations of the crust within the Wigner-Seitz approximation
\cite{Khan05,Sandulescu04}, valid for wavelengths smaller than the
radius of the Wigner-Seitz cell. Concerning the microscopic input for
the nuclear equation of state and the geometry of the structure, we
followed the work by Avancini et al.~\cite{Avancini09}.

Within this approach, we have calculated the excitation spectrum of a
periodic structure of parallel slabs, the lasagna phase. We have shown
that the structure can indeed induce non-negligible effects on the
excitation spectrum. In particular, we found that the sound velocity
of the usual acoustic mode depends on the direction of the
propagation, and, more surprisingly, that there is a second acoustic
mode whose dispersion relation is almost independent of $q_z$. In
addition, we found different optical branches, similar to the phonon
spectrum of ordinary crystals.

We have calculated the specific heat corresponding to this excitation
spectrum and found that its contribution is much more important than
that of individual neutrons, which is strongly suppressed due to the
superfluid gap $\Delta$. At temperatures relevant for neutron stars,
the main contributions to the specific heat come from the electrons
and from the acoustic collective modes. The latter cannot be obtained
within the Wigner-Seitz approximation. Due to the curious sound mode
whose energy is independent of $q_z$, the specific heat due to the
collective modes goes like $T^2$ instead of $T^3$. [With the same
  arguments, one would predict that in a rod structure (spaghetti
  phase), the specific heat should be linear in $T$.] However, it is
not clear whether this feature survives when the Coulomb interaction,
which has been neglected here, will be taken into account.

Of course, in order to treat the complex geometry, we were obliged to
make a couple of approximations. Contrary to the microscopic approaches
based on the Quasiparticle-Random-Phase-Approximation (QRPA)
\cite{Khan05,Sandulescu04}, we rely on the assumption that the modes
can be described hydrodynamically, which implies in particular that
the local neutron and proton Fermi surfaces stay spherical at all
times. This assumption is justified if all spatial variations are slow
compared to the superfluid coherence length and the temporal
variations are slow compared to the superfluid gap. Both assumptions
are not very well fulfilled. However, we have cited examples where
hydrodynamics gives reasonable answers even beyond these very
restrictive limits, and we believe that the results are at least
qualitatively correct. The most serious limitation of the present work
is probably that the Coulomb interaction has been neglected. The
Coulomb interaction between the protons results in an additional
coupling between neighboring cells, which can have important
consequences for the excitation spectrum. In the approaches of
Refs.~\cite{Pethick10, Cirigliano11}, it was accounted for by
including the elasticity of the Coulomb lattice. In our more
microscopic approach, the Coulomb potential would have to be included
from the beginning into the proton chemical potential
$\mu_p(\vek{r},t)$ in the Euler equation (\ref{eq:euler}). This is a
difficult task which will be left for future studies.

It has to be stressed that the contribution of the collective modes
studied here is potentially more important than other contributions,
notably the contribution from individual neutrons. Therefore it is
interesting to pursue their investigation and to include the
additional contribution to the specific heat in studies of neutron
star thermal evolution.

\acknowledgments 
We are indebted to C. Da Providencia for providing us
with the data for densities and geometries of the different pasta
phases in the model of Ref. \cite{Avancini09} as well as to M. Fortin for
providing us with the data of Ref.~\cite{Fortin09}. We thank
S. Chiacchiera, D. Pe\~na, and M. Fortin for discussions. This work
was supported by ANR (project NEXEN), and by CompStar, a research
networking programme of the European Science Foundation.

\bibliography{biblio}

\end{document}